\documentclass{aa}  

\usepackage{txfonts}
\usepackage[T1]{fontenc}
\usepackage{amsmath}
\usepackage{array,multirow,graphicx}
\DeclareMathAlphabet{\mathsfit}{T1}{\sfdefault}{\mddefault}{\sldefault}
\usepackage[dvipsnames]{xcolor}
\begin{document} 

   \title{Radiative acceleration calculation methods and abundance anomalies in Am stars}


   \author{Alain Hui-Bon-Hoa\inst{1}}

   \institute{IRAP, Universit\'e de Toulouse, CNRS, UPS, CNES, Toulouse, France\\
            \email{alain.hui-bon-hoa@univ-tlse3.fr}
 }

   \date{Received 26 August 2024; accepted 10 October 2024}

  \abstract
   {Atomic diffusion with radiative levitation is a major transport process to consider to explain abundance anomalies in Am stars. Radiative accelerations vary from one species to another, yielding different abundance anomalies at the stellar surface.}
   {Radiative accelerations can be computed using different methods: some evolution codes use an analytical approximation, while others calculate them from monochromatic opacities. We compared the abundance evolutions predicted using these various methods.}
   {Our models were computed with the Toulouse-Geneva evolution code, in which both an analytical approximation (the single-valued parameter method) and detailed calculations from Opacity Project (OP) atomic data are implemented for the calculation of radiative accelerations. The time evolutions of the surface abundances were computed using macroscopic transport processes that are able to reproduce observed Am star surface abundances in presence of atomic diffusion, namely an ad hoc turbulent model or a global mass loss.}
   {The radiative accelerations obtained with the various methods are globally in agreement for all the models below the helium convective zone, but can be much greater between the bottom of the hydrogen convective zone and that of helium. The time evolutions of the surface abundances mostly agree within the observational error, but the abundance of some elements can exceed this error for the least massive mass-loss model. The gain in computing time from using analytical approximations is significant compared to sequential calculations from monochromatic opacities for the turbulence models and for the least massive wind model; the gain is small otherwise. Test calculations of turbulence models with the tabulated OPAL opacities yield quite similar abundances as OP for most elements but in a much shorter time, meaning that determining Am star parameters can be done using a two-step method.}
   {}


   \maketitle
%

\section{Introduction}

The Am stars are characterised by surface abundance anomalies with typically underabundant calcium and/or scandium and an excess of Fe-peak and heavier elements \citep{Conti1970}. Some effects of these anomalies are visible when one tries to spectrally classify these stars, as \cite{Titus_Morgan1940} did in the Hyades open cluster. They were the first to define them as a separate group of stars. At the spectral classification level, Am stars are in the hydrogen spectral type interval A5-F2 \citep{Roman_etal1948}, so some authors call them AmFm. An extension towards earlier types was found via detailed atmosphere analyses, which showed that stars such as Sirius A \citep{Kohl1964} and 68 Tau \citep{Conti_etal1965} have Am star abundance anomalies. \cite{Preston1974} set the temperature range to the interval [7000;10000]~K, which corresponds to stellar masses of 1.5 to 3~$M_\odot$ \citep{Richer_etal2000}.

Following the idea of \cite{Michaud1970}, that radiatively driven atomic diffusion could account for the abundance anomalies observed in Ap stars, \cite{Watson1970, Watson1971} and \cite{Smith1971} investigated the possibility that the same transport process produces the anomalous abundances of Am stars. The migration of the chemical species via atomic diffusion is mainly due to the competition between two forces: gravity, which leads all the elements to settle towards the stellar centre, and the radiative accelerations ($g_\mathrm{rad}$), which drag them outwards. The latter result from the absorption of photons by each kind of ion, which then gain momentum from the radiation field. The calculation of $g_\mathrm{rad}$ therefore requires the knowledge of the way each species interacts with the radiation field, for each layer of the star. 

The first computations of $g_\mathrm{rad}$ in stellar envelopes were made by \cite{Watson1971} just below the base of the superficial convective zone (SCZ) of Am stars.  \cite{Kobayashi_Osaki1973} extended this work to slightly deeper regions, but for fewer elements. These studies considered only bound-bound transitions. \cite{Michaud_etal1976} provided approximate formulae based on experimental data to compute $g_\mathrm{rad}$ for a wider set of chemical elements and range of use in the envelope. They also considered absorptions via photoionisation.
 
The advent of extensive atomic databases, such as the Opacity Project \citep[OP; e.g.][]{Seaton2005} and OPAL \citep{Iglesias_Rogers1996}, has allowed radiative accelerations to be computed in detail for many elements in the whole envelope. The first evolution code to implement such a method was the Montpellier-Montr\'eal code \citep{Richer_etal1998}, which uses OPAL data. It has been used to investigate Am stars \citep{Richer_etal2000,Richard_etal2001,Michaud_etal2005,Vick_etal2010,Michaud_etal2011b}, among other kinds of stars. Calculations with OP data have been implemented in the MESA code \citep{Paxton_etal2011,Paxton_etal2013,Paxton_etal2015,Paxton_etal2018, Paxton_etal2019,Jermyn_etal2023}, first with the \cite{Hu_etal2011} method. This version of MESA has been used to constrain the turbulent mixing inside F and G stars  \citep{Moedas_etal2022,Moedas_etal2024}. Recently, \cite{Mombarg_etal2020} and \cite{Mombarg_etal2022} optimised the computation of $g_\mathrm{rad}$ in MESA to produce models used to predict gravity mode properties in $\gamma$ Dor stars.

However, computations of $g_\mathrm{rad}$ from monochromatic opacities require a great deal of computing time due to numerous integrations over frequencies. As an alternative, analytical formulae have been developed \citep[][and references therein]{LeBlanc_Alecian2004}, leading to the so-called single-valued parameter (SVP) method. This method significantly reduces computing time and has been implemented in the Toulouse-Geneva evolution code \citep{Theado_etal2009} and in CESAM2k20 \citep{Deal_etal2018}. It also allows elements that are missing in the OP set to be considered, such as scandium \citep{Hui-Bon-Hoa_etal2022b}. \cite{Alecian_LeBlanc2020} recently improved their method, which is now implemented in the Toulouse-Geneva evolution code (TGEC), along with the detailed calculation of $g_\mathrm{rad}$ from OP monochromatic cross-sections.

In this paper we compare the results of these different methods of $g_\mathrm{rad}$ calculation and their effect on the computed evolution of surface abundances of Am stars. After recalling the physics used in TGEC (Sect.~\ref{tgec}), we detail the implementation of the various $g_\mathrm{rad}$ calculation methods we used (Sect.~\ref{g_rad}). Section~\ref{results} compares the results of the different kinds of computations, and we discuss the context where analytic approximations could be valuable (Sect.~\ref{discussion}).


\section{Stellar models}\label{tgec}

\subsection{Basic input physics}

TGEC is a one-dimensional evolution code \citep{Hui-Bon-Hoa2008} in which the stellar structure is converged assuming hydrostatic equilibrium. In the latest version, the Rosseland mean opacities are computed dynamically with the OP monochromatic opacities from the OPCD v.3.3 data \citep{Seaton2005}, using the method described in \cite{Hui-Bon-Hoa2021} to save computing time. In this optimisation strategy, an opacity table is computed with the initial mixture. Values from this table are used when the relative difference between the abundances of the layer of interest and the initial value is smaller than a threshold, which we set to $10^{-4}$. If the relative abundance difference exceeds this threshold, the Rosseland mean is recalculated on the fly. If neighbouring layers have the same abundances (e.g. in a convective zone), only values of the OP grid points that have not been computed so far are calculated. Such an approach enables total consistency between the Rosseland opacity and the chemical mixture in all parts of the star, in particular because of the abundance variations induced by atomic diffusion. The nuclear reaction rates are from the NACRE compilation \citep{Angulo_etal1999}, and we used the OPAL2001 equation of state \citep{Rogers_Nayfonov2002}.\\

The macroscopic processes were chosen to best reproduce the observed abundances of Am stars, that is, either using a turbulence model without mass loss, as in \cite{Richer_etal2000}, or considering mass loss with convection as the only mixing process \citep{Vick_etal2010}. In the first case, a full and instantaneous mixing acts down to the Z-bump ($\log T=5.3$), hereafter referred to the superficial mixing zone (SMZ), and turbulent mixing is introduced below through a diffusion coefficient, $D_\mathrm{T}$, computed according to Eq.~1 of \cite{Richer_etal2000}, \begin{equation}
D_\mathrm{T}=\omega D(\mathrm{He})_0\left(\frac{\rho_0}{\rho}\right)^n\label{equation:RMT}
,\end{equation}
where $D(\mathrm{He})_0$ is the diffusion coefficient of helium at the density $\rho_0$. We chose to set the free parameters as in their `R1k-2' models, which reproduced the surface abundances of most of the Am stars they considered. Namely,  $\rho_0$  was set to $8\times10^{-6}\mathrm{g.cm^{-3}}$, $n$ to 2, and $\omega$ to $10^3$.

Models with mass loss see their outermost layers removed at each evolution time step \citep{Hui-Bon-Hoa_etal2022b}. If the innermost layer removed is deeper than the base of the SCZ, the evolution time step is adjusted to limit the layers removed to this convective zone, thus avoiding layers where atomic diffusion has yielded abundance inhomogeneities. We therefore considered chemically unseparated winds where the amount of each element lost at each time step only depends on its content in the removed layers. We adopted a mass-loss rate of $3.1\times10^{-14}~M_\odot/\mathrm {yr}$, which is logarithmically in the middle of the interval of values that matched most of the Am star abundances quoted by \cite{Vick_etal2010}.

\begin{figure}
   \centering
   \includegraphics[width=0.5\textwidth]{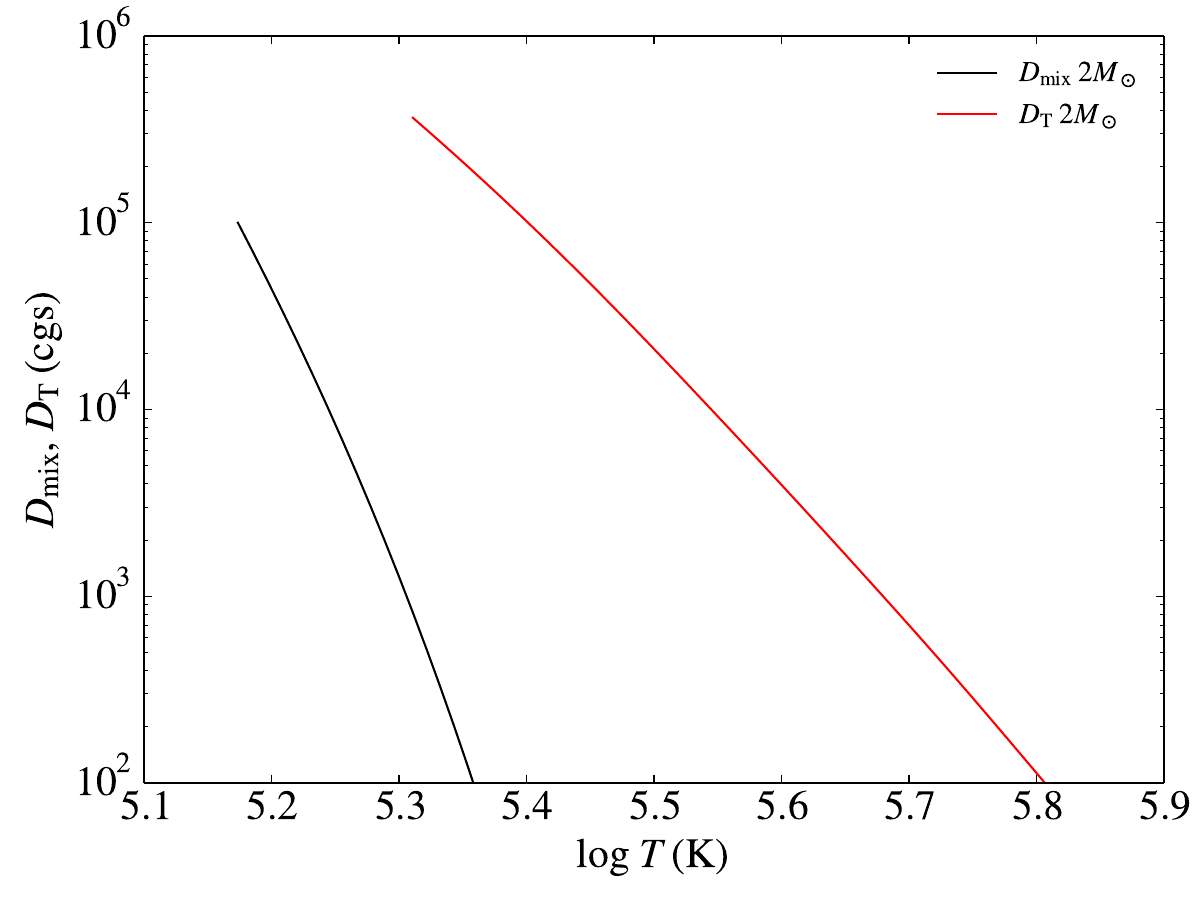}
      \caption{Mixing diffusion coefficients $D_\mathrm{mix}$ (black line) and $D_\mathrm{T}$ (red line) vs $\log T$ at mid-MS for a $2.0~{M_\odot}$ model without diffusion. $D_\mathrm{T}$ is only computed down to $\log T=5.3$ and $D_\mathrm{mix}$ down to the bottom of the SCZ, the layers above being mixed instantaneously.}
       \label{figure:codif}
\end{figure}

Convective zones are treated with the mixing-length theory \citep{Bohm-Vitense1958} with a mixing-length parameter ($\alpha=L/H_\mathrm{P}$) of 1.8. $L$ is the mixing length and $H_\mathrm{P}$ the pressure scale height. This value is set through a solar calibration \citep{Richard1999}.
The convective zones due to H, HeI, and HeII are assumed to be all connected such that the superficial mixing encompasses the layers between the top of the H convective zone down to the bottom of that of HeII, when present. A mild turbulent mixing is set at the bottom of the SCZ to avoid unrealistic discontinuities of the abundances at this boundary \citep{Theado_etal2009}. This mixing is parametrised by a diffusion coefficient, $D_\mathrm{mix}$, expressed as
\begin{equation}\label{mildMixing}
D_\mathrm{mix}=D_\mathrm{bcz}\exp\left(\frac{r-r_\mathrm{bcz}}{\Delta}\ln 2\right)
,\end{equation}
where $r$ is the current radius and $D_\mathrm{bcz}$ the diffusion coefficient at the bottom of the SCZ (at radius $r_\mathrm{bcz}$), set to $10^5 \mathrm{cm^2.s^{-1}}$. The depth of the mixing is controlled by $\Delta$, whose value is equal to 0.2~\% of the stellar radius. \cite{Freytag_etal1996} suggested values about ten times lower in their overshoot prescription for A-type stars, but we find that reducing $\Delta$ increased numerical noise in the abundance evolution, whereas the abundance change remained smaller than the observational uncertainty. The variations in $D_\mathrm{mix}$ and $D_\mathrm{T}$ versus temperature are shown in Fig.~\ref{figure:codif} for a $2.0~{M_\odot}$ model. 
We assumed that rotation is slow enough to allow us to neglect the mixing associated with meridional circulation.

The initial abundances we used are those prescribed by \cite{Serenelli2010} for the solar chemical composition, namely the meteoritic values of \cite{Lodders_Palme_Gail2009} for the refractory elements, and the photospheric values of \cite{Asplund_etal2009} for the remaining species. The metallicity $Z$ is equal to 0.0134.

\subsection{Atomic diffusion}

Atomic diffusion was implemented via the \cite{Chapman_Cowling1970} formalism, in which the various species move with respect to the major component, which is hydrogen in the envelope of main sequence (MS) stars. Diffusion of helium was treated using the \cite{Montmerle_Michaud1976} method, and we used the values of \cite{Paquette_etal1986} for the diffusion coefficients. Fourteen chemical species were considered: H, He, C, N, O, Ne, Na, Mg, Al, Si, S, Ar, Ca, and Fe. The set of metals was constrained by the data available to compute $g_\mathrm{rad}$, and we only included the 12 species in common for the different methods we used. The $g_\mathrm{rad}$ were computed at each diffusion time step according to the local abundances.


\section{Radiative accelerations}\label{g_rad}

\subsection{Computation from monochromatic cross-sections}

The radiative accelerations express the momentum transfer from the radiation field to the various ions of the stellar medium. For atomic species $i$:
\begin{equation}
g_\mathrm{rad}(i)=\frac{1}{c}\frac{1}{M_i}\int \sigma_\nu^\mathrm{mta}(i)\mathcal{F}_\nu\mathrm{d}\nu
,\end{equation}
with $c$ being the speed of light, $M_i$ the mass of atom $i$, $\sigma_\nu^\mathrm{mta}(i)$ the cross-section expressing the momentum transferred to the atom at frequency $\nu$, and $\mathcal{F}_\nu$ the monochromatic radiation flux. The strength of this transfer is closely related to the monochromatic absorption cross-sections, $\sigma_\nu(i),$ used to compute Rosseland mean opacities. However, $\sigma_\nu(i)$ cannot be directly used in place of $\sigma_\nu^\mathrm{mta}(i),$ firstly because scattering does not contribute to $g_\mathrm{rad}$. Additionally, several physical processes have to be considered, such as the averaging over the different ionisation stages of a given chemical species and the momentum sharing between electrons and atoms in the case of bound-free or free-free absorptions, as discussed by \citet[hereafter S97]{Seaton1997} and \cite{Richer_etal1998}.

The monochromatic opacities excerpted from the OP data do not include scattering, which was therefore computed with the actual average number of electrons per ion of the mixture. As for the above-mentioned physical processes, following S97, only the momentum sharing occurring during photoionisation was considered; the choice of the weighting method for the averaging between ions should  have little influence on the transport of the chemicals in the temperature domain of Am stars because of the thickness of their SCZ. The adopted weighting is then the same as for opacity calculations, allowing one to write $\sigma_\nu^\mathrm{mta}(i)=\sigma_\nu(i)-\sigma_\nu^\mathrm{mte}(i)$ (Eq.~31 of S97), where $\sigma_\nu^\mathrm{mte}(i)$ expresses the momentum transferred to the electrons of species $i$ and is provided in the OPCD files.

The implementation of $g_\mathrm{rad}$ calculations in stellar interiors makes use of the diffusion approximation \citep{Milne1927} to evaluate the radiative flux since the medium is optically thick. Defining $f_\nu=(\mathrm{d}B_\nu/\mathrm{d}T)/(\mathrm{d}B/\mathrm{d}T)$, we then used an expression close to those of \cite{Seaton2005} and \cite{Seaton2007}, writing $g_\mathrm{rad}(i)$ as
\begin{equation}
g_\mathrm{rad}(i)=\frac{1}{c}\frac{\mu}{\mu_i}\mathcal{F}\kappa_\mathrm{R}\gamma_i
,\end{equation}
where $\mu$ and $\mu_i$ are the mean atomic mass and the atomic mass of species $i,$ respectively,
\begin{equation}
\gamma_i=\int \frac{\sigma_\nu^\mathrm{mta}(i)}{\sigma_\nu}f_\nu\mathrm{d}\nu\label{equation:gamma}
 ,\end{equation}
 and the integrated flux, $\mathcal{F}$, 
\begin{equation}
\mathcal{F}=\pi B(T_\mathrm{eff})\left(\frac{R_*}{r}\right)^2
 ,\end{equation}
where $B(T_\mathrm{eff})$ is the value of the Planck function at the effective temperature, $T_\mathrm{eff}$, and $r$ and $R_*$ are the current and stellar radii, respectively.

In each layer of the stellar structure, $g_\mathrm{rad}(i)$ was computed in the same way as in the routines available in the OPCD: a set of 16 grid points (4 temperatures, $T,$ by 4 electronic densities, $N_\mathrm{e}$) was selected around the ($T$, $N_\mathrm{e}$) of the current layer, providing 16 $\gamma_i$ values. A bi-cubic interpolation of the logarithms of the $\gamma_i$ was then performed to determine its value for the layer of concern. As atomic diffusion leads to abundance variations for all the elements at a given time, we computed the $\gamma_i$ with the current mixture instead of using the abundance variation factor for an individual element, $\chi$, introduced by \cite{Seaton1997}. A similar implementation was used by \cite{Hu_etal2011} in their study of sdB stars and has been implemented in the stellar evolution code MESA \citep{Paxton_etal2015,Paxton_etal2019}.

Due to the momentum transfer to the electron during photoionisation, $g_\mathrm{rad}$ can be negative in some rare cases, especially for Na around $\log T=4.4$. In such situations, the {\tt ax.f} routine of the OPCD sets $\log\gamma_i$ to -30 at the OP grid points before interpolation. Using this method, the positive values of $g_\mathrm{rad}(\mathrm{Na})$ we obtained differ by up to several orders of magnitude from the computations at the nearest OP grid points in these layers. This problem was circumvented by using a bi-linear interpolation of the four central $\log \gamma_i$ values around the  ($T$, $N_\mathrm{e}$) of interest when any of the outer 12 values were negative. If negative values were present in the central square, the bi-linear interpolation was done with the $\gamma_i$ instead of their logarithms.

For a given stellar structure, the evaluation of integral (\ref{equation:gamma}) is the most time-consuming part of the $g_\mathrm{rad}$ calculation. To reduce wall-time computing time, we took advantage of the strategy applied by \cite{Hui-Bon-Hoa2021} for the computation of Rosseland opacities: along with the Rosseland means, all the values of $\gamma_i$ for all species $i$ were computed at the beginning of the evolution with the initial abundances set at each of the grid points of the OP temperature-electronic density. If the abundances departed from the initial mixture in a given layer of the model, new values of $\gamma_i$ were computed accordingly. Tests performed with three masses well distributed in the Am star mass interval (namely 1.5, 2, and 2.5~$M_\odot$) allowed the optimal threshold for the relative abundance change to be set to $5 \times 10^{-2}$. Higher values led to different surface abundance evolutions, whereas lower ones only increased computing time without significantly changing the surface abundance evolution. 

\cite{Mombarg_etal2022} implemented  a further step of computational optimisation in MESA using precomputed $g_\mathrm{rad}$ tables in their modelling of $\gamma$ Dor stars, but they considered a radiative envelope divided into two parts with averaged abundances in each to compute $g_\mathrm{rad}$. This approximation cannot be used here since the abundance variations in the radiative envelope of Am stars \citep[e.g.][]{Richer_etal2000,Vick_etal2010} are much greater than those present in $\gamma$ Dor stars \citep{Mombarg_etal2020}. 

\subsection{Analytical approximations}

In stellar interiors, where the radiative flux can be estimated via diffusion approximation, analytical formulae have been developed to compute $g_\mathrm{rad}$. Before the availability of large atomic data databases, atomic data were lacking for many of the ionisation stages present in stellar envelopes \citep[e.g.][]{Alecian_LeBlanc2020}. After the pioneering work of \cite{Michaud_etal1976}, the first theoretical developments of the now-called SVP method were devised by \cite{Alecian1985} and \cite{Alecian_Artru1990a}. The method has been improved gradually \citep{Alecian_LeBlanc2000,Alecian_LeBlanc2002,LeBlanc_Alecian2004,Alecian_LeBlanc2004,Alecian_LeBlanc2020} and consists of two analytical expressions -- one for bound-bound transitions and one for photoionisation -- whose coefficients are tabulated. Six parameters are needed per ionisation stage of each chemical species, four for the line absorptions and two for bound-free transitions. Three of these quantities were computed from the OP atomic data available through the TOPbase\footnote{https://cds.unistra.fr/topbase/topbase.html} database \citep{Cunto_etal1993}, and the others were obtained by fitting the $g_\mathrm{rad}$ to those of \cite{Seaton2007}. We refer the reader to \cite{Alecian_LeBlanc2020} for the details regarding their determination. These data were computed using  solar-metallicity stellar models at mid-MS so that they can be used for most MS population I stars. Tables are available for stellar masses between 1 and 5~$M_\odot$ with a 1~$M_\odot$ step \citep{Alecian_LeBlanc2004} or in the interval [1;10]~$M_\odot$ with a variable step \citep{Alecian_LeBlanc2020}. The coefficients were interpolated if the required mass was between values of the mass grid. The accuracy of the parameters derived from atomic data is better in  \cite{Alecian_LeBlanc2020} than in \cite{Alecian_LeBlanc2004}. The fitting method of the free parameters has also been improved.

Since the $g_\mathrm{rad}$ of the different chemical species were calculated independently, the effect of an abundance change of a given element on the $g_\mathrm{rad}$ of the others, as discussed by \cite{Richer_etal1998}, was ignored. The SVP2020 data are publicly available\footnote{https://voparis-gradsvp.obspm.fr/}.

\begin{figure*}
   \centering
   \includegraphics[width=\textwidth]{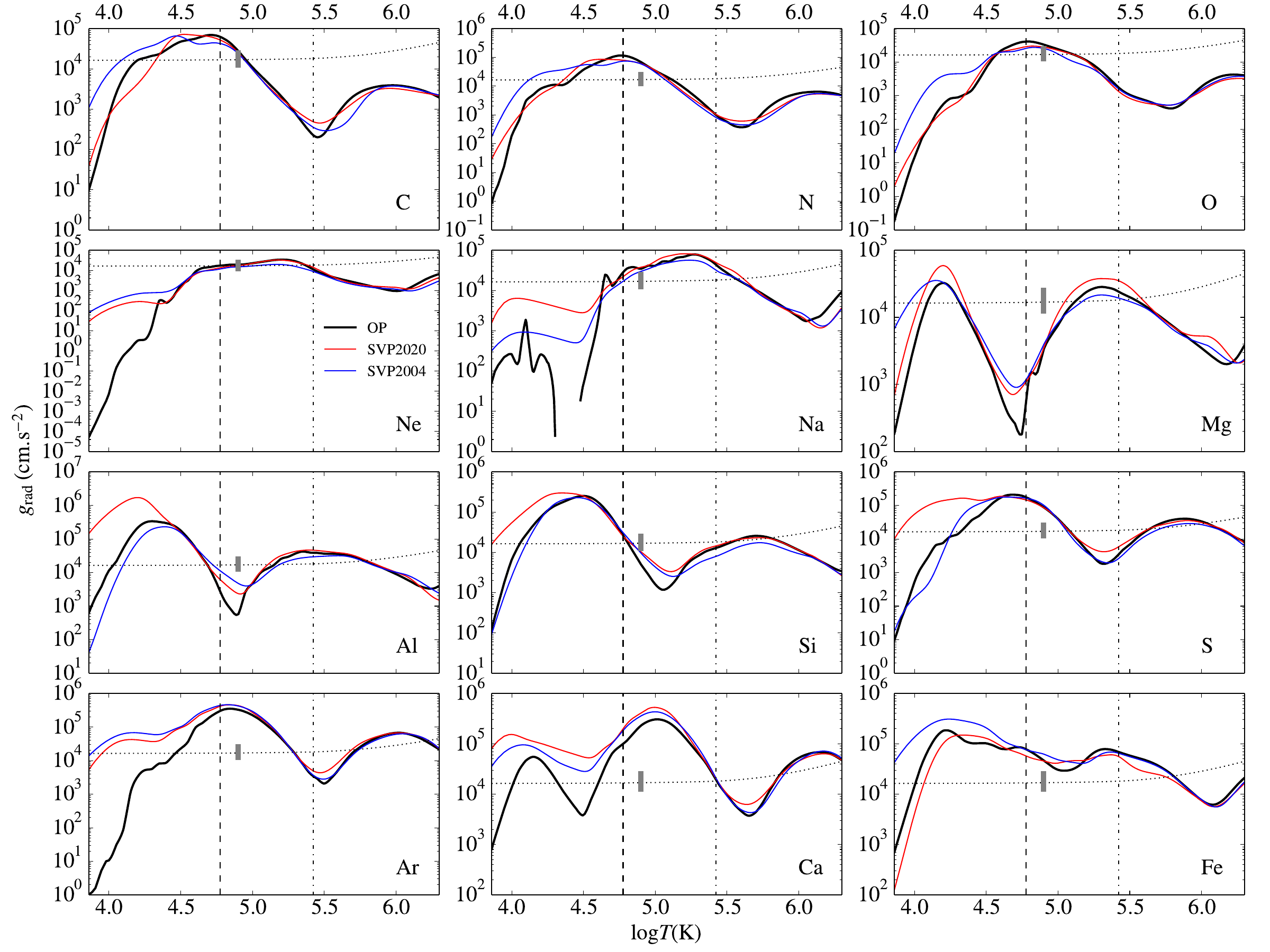}
      \caption{Radiative accelerations vs $\log T$ at mid-MS computed with OP data (thick black lines) and the SVP2020 and SVP2004 methods (red and blue lines, respectively) for a chemically homogeneous $1.5~{M_\odot}$ model. The dotted lines denote the local gravity (in absolute values). The vertical lines show the location of the bottom of the H and He convective zones (dashed and dash-dotted lines, respectively). The grey bar near the centre of each panel illustrates the maximum acceptable difference in the $g_\mathrm{rad}$ (see the main text). The gap in the OP curve for Na corresponds to negative values (not shown).
      }
         \label{figure:g_rad1.5}
\end{figure*}

\begin{figure*}
   \centering
   \includegraphics[width=\textwidth]{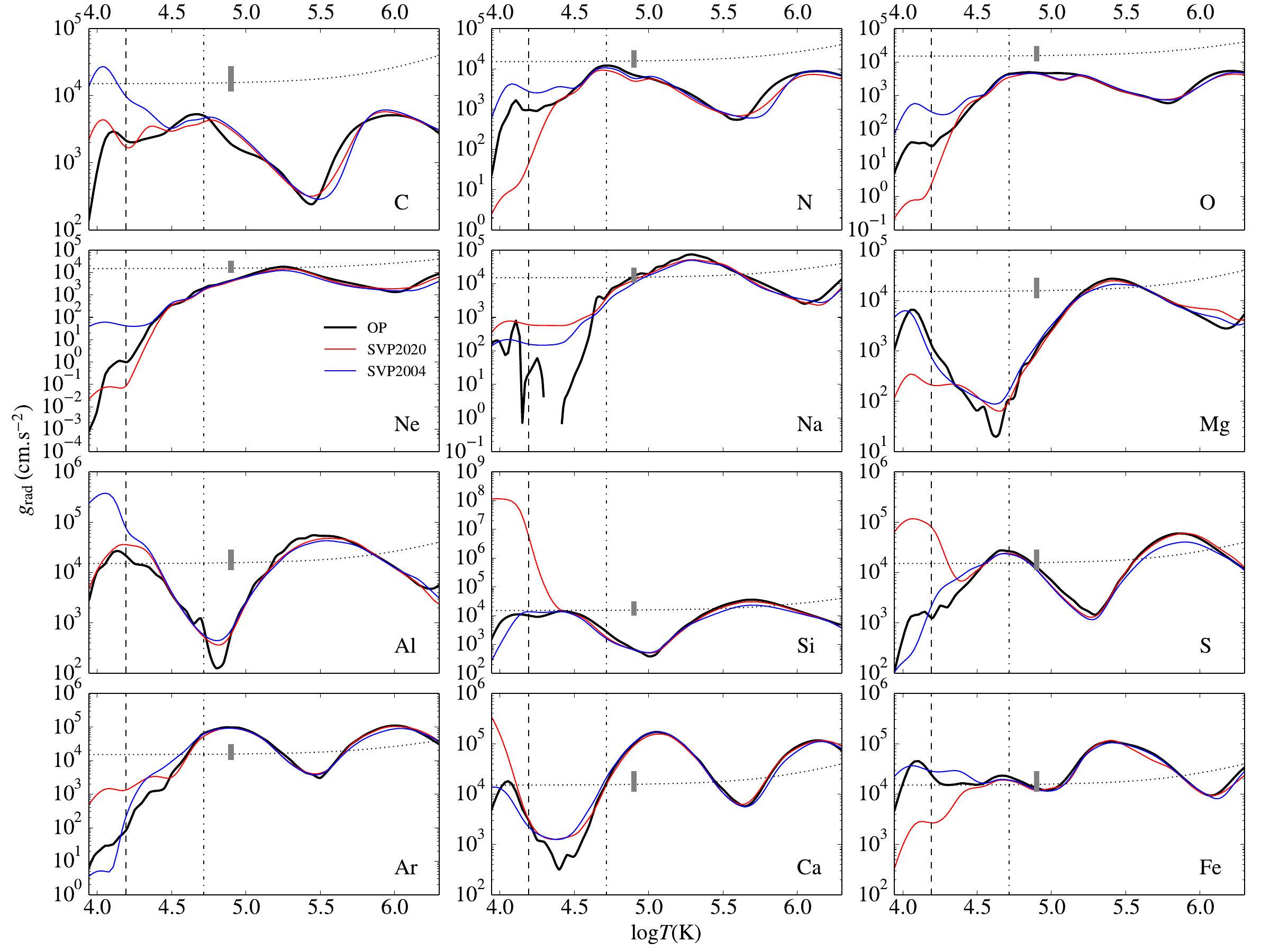}
      \caption{Same as Fig.~\ref{figure:g_rad1.5} but for the 2~$M_\odot$ model.
              }
         \label{figure:g_rad2.0}
\end{figure*}

\begin{figure*}
   \centering
   \includegraphics[width=\textwidth]{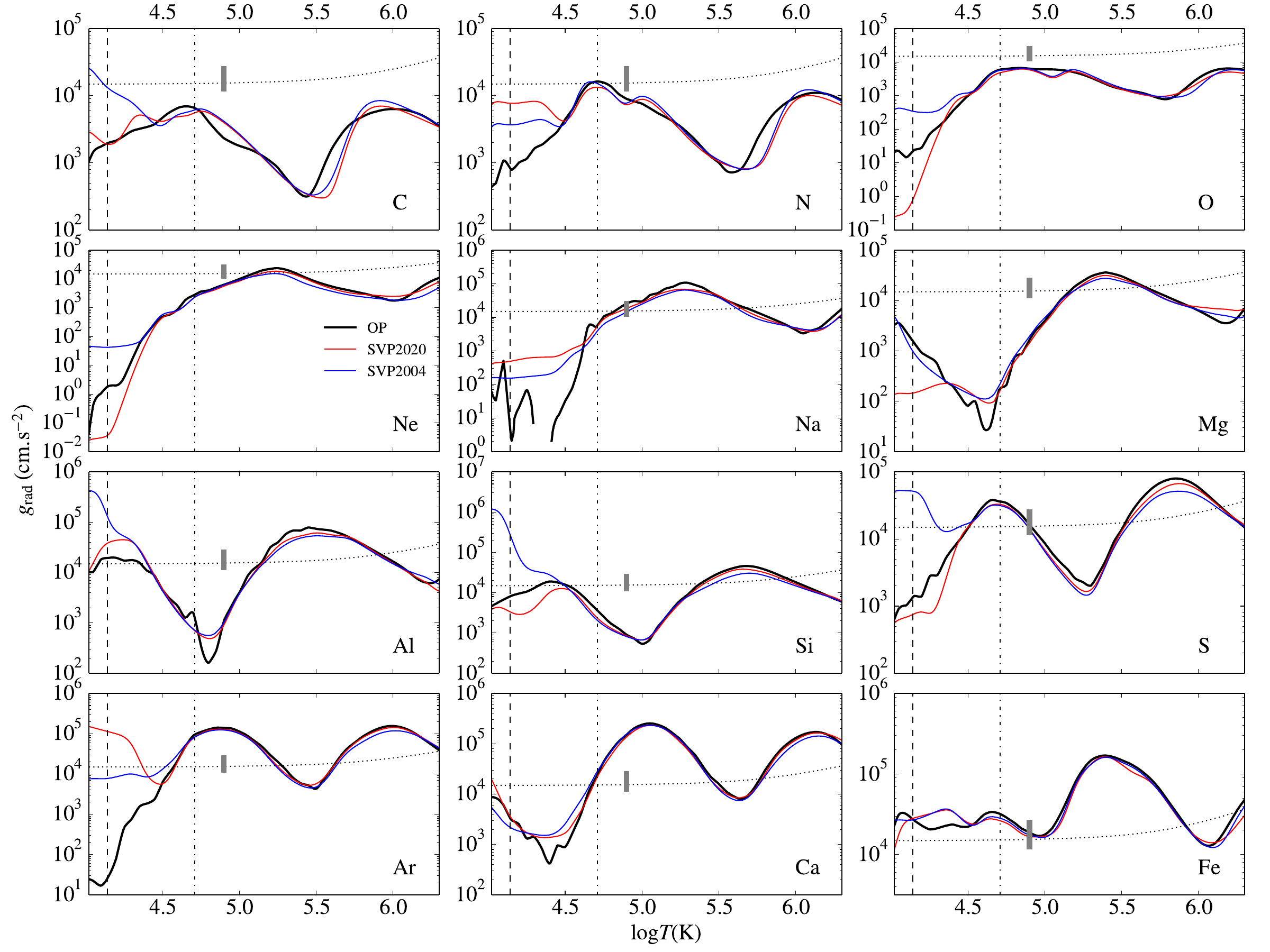}
      \caption{Same as Fig.~\ref{figure:g_rad1.5} but for the 2.5~$M_\odot$ model.
              }
         \label{figure:g_rad2.5}
\end{figure*}


\section{Results}\label{results}

Though \cite{Richer_etal2000} used the mass interval [1.5;3]~$M_\odot$ for Am stars, we set the upper mass boundary to 2.5~$M_\odot$ because the effective temperature of a 3~$M_\odot$ star enters the Am star temperature domain only at the end of the MS, whereas a 2.5~$M_\odot$ star spends the second half of its MS life below 10.000~K. We thus considered the three masses 1.5, 2, and 2.5~$M_\odot$ to model Am stars. 

\subsection{Comparison of the various $g_\mathrm{rad}$ calculation methods}

In this section we compare the $g_\mathrm{rad}$ computed using the two versions of the SVP approximation and the results obtained with the detailed calculation using OP monochromatic data. Figure~\ref{figure:g_rad1.5} shows the $g_\mathrm{rad}$ of the various elements we considered versus temperature for the $1.5~{M_\odot}$ model at mid-MS with homogeneous solar abundances. Being on a mass grid point of the SVP2004 dataset, this mass is between two SVP2020 grid points. The interpolation method used in the SVP codes led to negative $g_\mathrm{rad}$ values for N and Si with SVP2020 at places where they should be positive. This issue was circumvented by using a linear interpolation instead. Improvements to the interpolation method will be provided in a forthcoming release of the SVP data and codes (G. Alecian, priv. comm.).

Since the SVP method is an approximation of the more detailed OP calculations, \cite{Alecian_LeBlanc2020} defined an acceptable deviation between the $g_\mathrm{rad}$ yielded by SVP and OP of $0.3~\mathrm{dex}$. This  value is not related to observations, but to knowledge regarding atomic physics. Below the hydrogen convective zone, the differences between each of the two versions of SVP and the OP values are mostly within this limit for the $1.5~{M_\odot}$ model. The differences can be much larger closer to the stellar surface, but they will not have any consequence on the abundance evolution there since we assume that the mixing in the SCZ is efficient enough to hinder the appearance of any chemical inhomogeneity. However, for some elements (C, Al, and Si), some discrepancies are slightly above the deviation limit; however, this occurs when $g_\mathrm{rad}$ is weak compared to gravity and therefore should not have any significant effect since the motion of the chemicals is dominated by gravity at this point. The two SVP versions yield $g_\mathrm{rad}$ values in good agreement with each other below the hydrogen convective zone. The agreement between the three methods is even better below the He convection zone, but this mixing zone only exists when He is abundant enough. We can therefore expect a much more similar surface abundance evolution for models where He gravitational settling is slow enough. We note that the differences between the SVP and OP $g_\mathrm{rad}$ can be larger in this study compared to the Alecian \& LeBlanc series of papers because, unlike them, we did not apply the smoothing method used in the OPserver code. This smoothing is not suitable when dealing with a depth-dependent composition \citep{Seaton2005}, and our $g_\mathrm{rad}$ curves can show some extra bumps. For Na, the $g_\mathrm{rad}$ computed with OP data can be negative (i.e. directed towards the centre) around $\log T\simeq 4.4$, which cannot be reproduced by the SVP method. In any case, this happens in the H convective zone for this stellar mass and is therefore of no consequence.

Results for the 2 and 2.5~$M_\odot$ models are presented in Figs.~\ref{figure:g_rad2.0} and \ref{figure:g_rad2.5}, respectively. The differences between the SVP and OP $g_\mathrm{rad}$ values are mostly smaller than the deviation limit below the He convective zone, apart from locations where $g_\mathrm{rad}$ is much weaker than gravity. The situation worsens when we consider upper layers: we find deviations much greater than the limit below the H convection zone for almost all the elements. A large impact on the abundance stratification of Si and S near the stellar surface is expected when SVP2020 is used, since their $g_\mathrm{rad}$ are greater by several orders of magnitude than those computed with SVP2004 or OP just below the H convection zone. For some other elements (O, Ne, Na, and Mg), the impact should be small despite the difference in $g_\mathrm{rad}$ since it is much weaker than gravity there for both masses. This is also the case for N and Ar in the $2~{M_\odot}$ model. The situation is less clear for carbon, whose SVP2004 $g_\mathrm{rad}$ for the $2~{M_\odot}$ model and SVP2020 $g_\mathrm{rad}$ for the $2.5~{M_\odot}$ model are much closer to gravity compared to the OP or to the other SVP version computation. This is also the case for N with SVP2020 in the $2.5~{M_\odot}$ model, albeit to a lesser extent. For Al, we find discrepancies beyond the acceptable limit just below the H convective zone with SVP2004 for the $2~{M_\odot}$ model and with SVP2020 for the $2.5~{M_\odot}$ model. The difference between the iron SVP2020 $g_\mathrm{rad}$ and those from OP is also greater than the limit just below the H convective zone for the $2~{M_\odot}$ model, whereas the three calculations agree within this limit for the $2.5~{M_\odot}$ model. The one element whose $g_\mathrm{rad}$ deviations are below the limit for both models is Ca, though not at the dip around $\log T=4.5$, but this has a negligible impact since $g_\mathrm{rad}$ is much weaker than gravity there.

\subsection{Effect on the surface abundance evolutions of Am stars}

We compared the time evolution of model Am stars surface abundances whose $g_\mathrm{rad}$ were computed with each of the three abovementioned methods. These models were evolved with the macroscopic transport prescriptions that provided the best agreement with observed abundances up to now, namely the \cite{Richer_etal2000} turbulence model (hereafter RMT mixing) and the \cite{Vick_etal2010} wind model\footnote{The models computed for the surface abundance evolutions are available at https://doi.org/10.5281/zenodo.13905300}. Other macroscopic transport processes can be investigated, but this is beyond the scope of this paper. 

\begin{figure*}
   \centering
    \includegraphics[width=\textwidth]{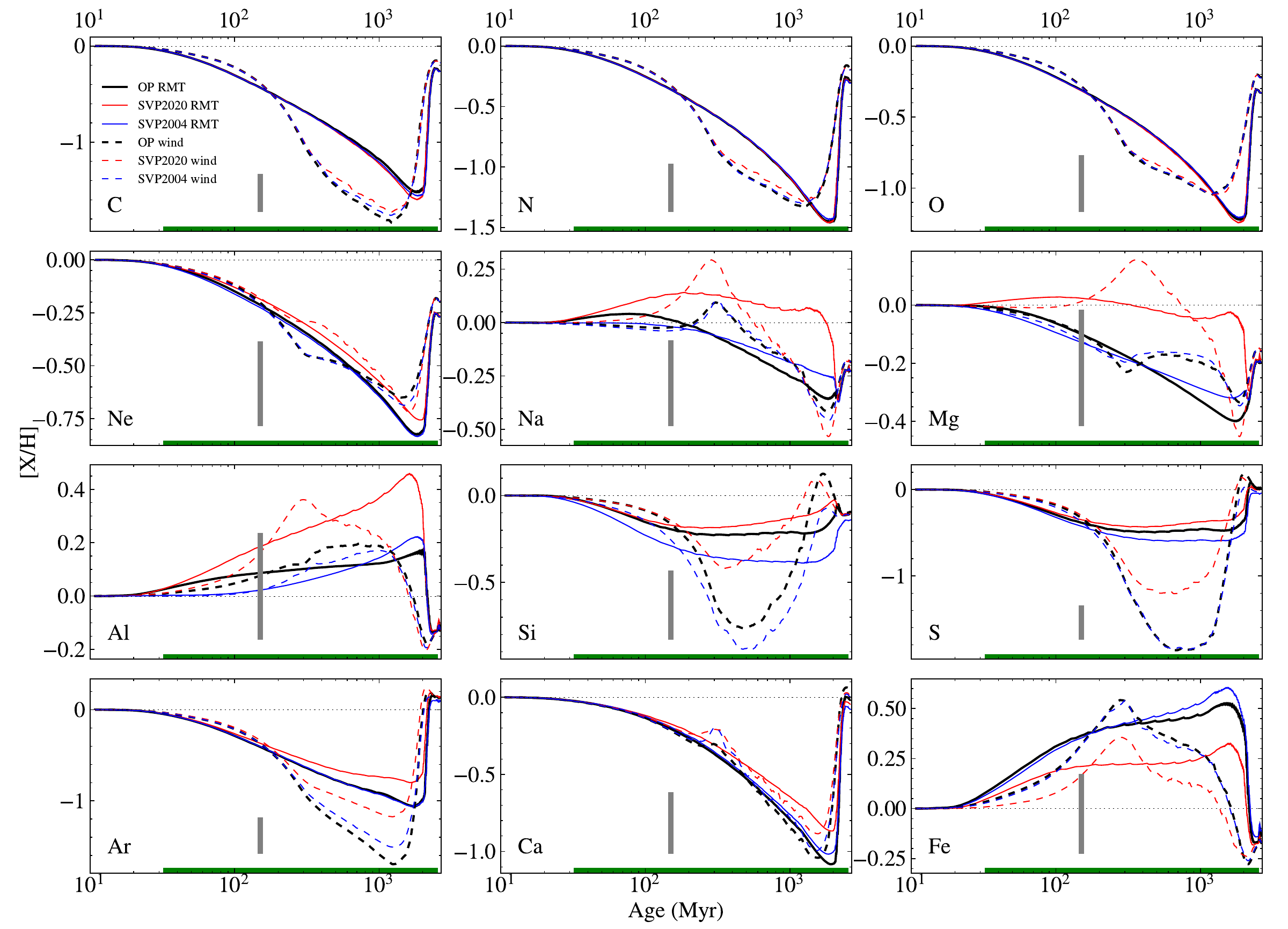}
      \caption{Time evolution of the surface abundances for the 1.5~${M_\odot}$ model with $g_\mathrm{rad}$ computed with OP (thick black line), SVP2020 (red line), and SVP2004 (blue line), expressed as the difference between the logarithm of the current ratio to H in number fractions and the corresponding initial value. Solid lines represent the models with RMT mixing, and dashed lines the mass loss (`wind') models. The thin dotted line represents the initial abundance. In each panel, the grey bar shows the typical abundance uncertainties (see the main text). The horizontal green bar represents the MS phase.
              }
         \label{figure:abSurf1.5}
\end{figure*}

Figure~\ref{figure:abSurf1.5} shows the surface abundance time evolution of the various elements of this study for the $1.5~M_\odot$ model, either when RMT mixing is considered or when mass loss is taken into account. If we adopt a conservative abundance determination uncertainty of $\pm 0.2$~dex \citep[e.g.][]{Gebran_etal2010}, the three methods agree with each other for the RMT mixing models. This is also true for the wind models for most elements. In the case of Si, the abundances computed using both SVP versions and OP agree within the uncertainties, but the difference between SVP2020 and SVP2004 is larger. For Mg, S, and Ar, the SVP2020 abundance track differs strongly from that of OP. The abundance discrepancies between the calculation methods can be partly explained by the $g_\mathrm{rad}$ differences where chemical separation occurs (see Fig.~\ref{figure:g_rad1.5}), that is, below $\log T=5.3$ -- or the base of the SCZ if deeper -- for the RMT mixing models. For the wind models, the bottom of the He convective zone is slightly closer to the surface than shown in Fig.~\ref{figure:g_rad1.5} because of mass loss, and chemical separation happens below a temperature $T_\mathrm{sep}\simeq 10^{5.3}$, with $T_\mathrm{sep}$ decreasing slowly with time. When the He abundance is too low to maintain a convective zone, around 300~Myr, $T_\mathrm{sep}$ drops to $\log T_\mathrm{sep}\simeq4.8$, which is the lowest temperature at which separation occurs during the MS life; the SCZ then thickens again due to evolution. This explains the extrema of the abundances of several elements. Another reason for the abundance differences is the variation in the sensitivity of $g_\mathrm{rad}$ to an abundance change from one element to the other, as illustrated, for instance, by the difference in the behaviour of the abundances of C and Ar despite similar $g_\mathrm{rad}$ dependences with depth. Additionally, for many elements, wind models exhibit larger abundance spreads because $T_\mathrm{sep}$ varies during the evolution, whereas separation occurs at a constant temperature for the RMT mixing models. As a general trend, the SVP2004 abundances are closer to the OP calculation than SVP2020.

\begin{figure*}
   \centering
    \includegraphics[width=\textwidth]{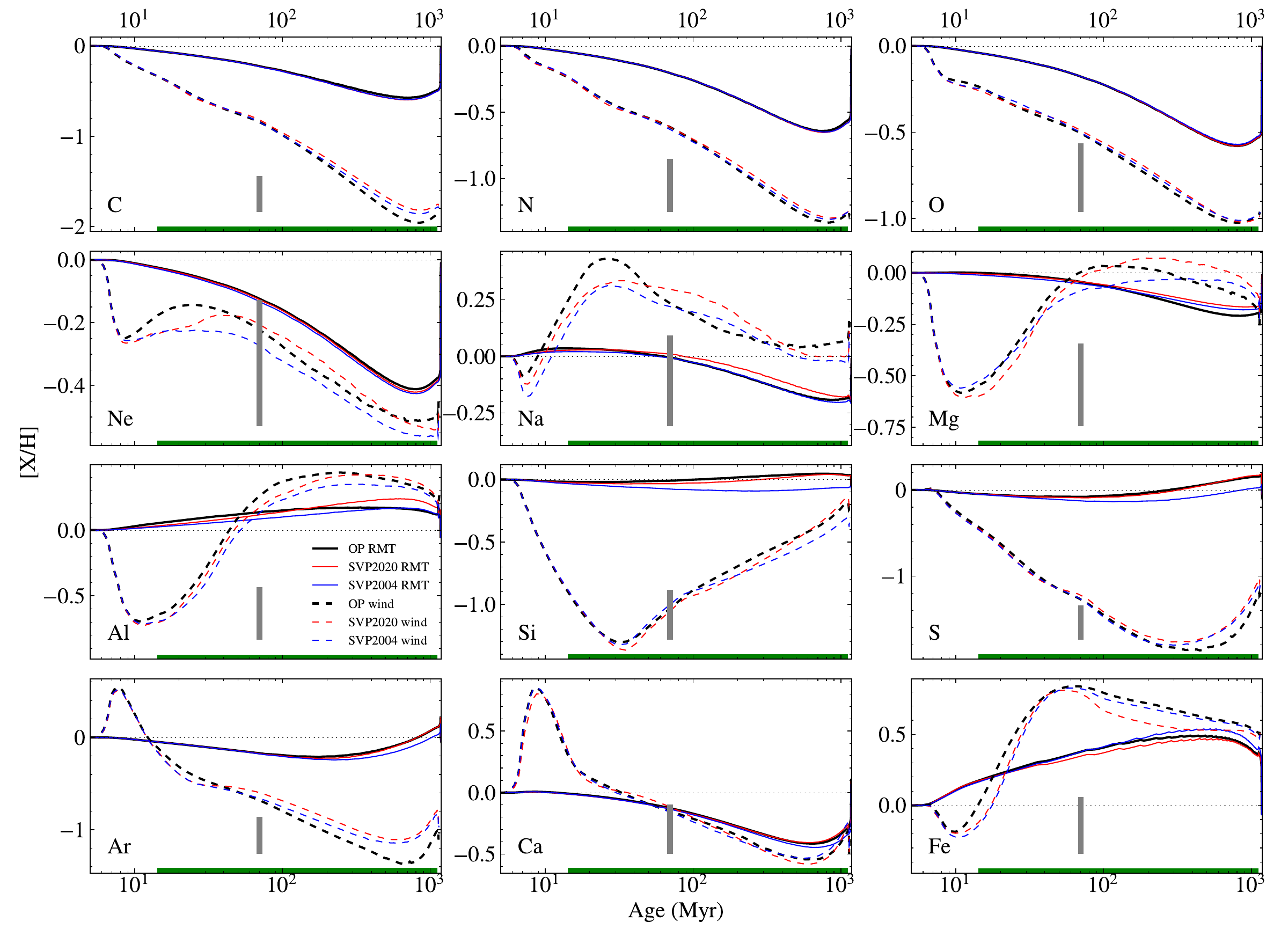}
      \caption{Same as Fig.~\ref{figure:abSurf1.5} but for the 2~$M_\odot$ models.
              }
         \label{figure:abSurf2.0}
\end{figure*}

\begin{figure*}
   \centering
    \includegraphics[width=\textwidth]{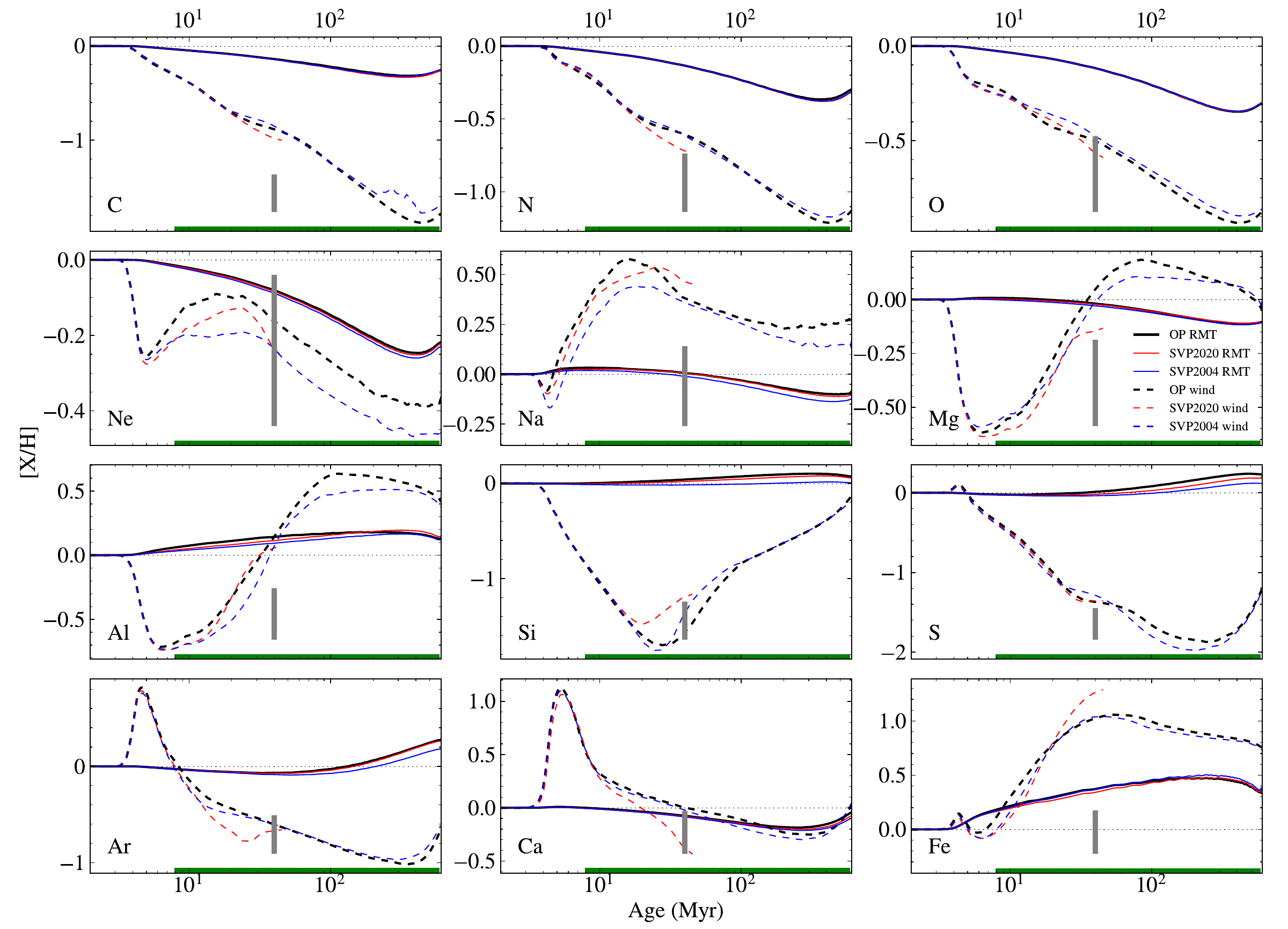}
      \caption{Same as Fig.~\ref{figure:abSurf1.5} but for the 2.5~$M_\odot$ models. The SVP2020 wind model stopped because the abundance of Fe was too strong at the Z-bump (see the main text).
              }
         \label{figure:abSurf2.5}
\end{figure*}

Concerning the 2 and $2.5~{M_\odot}$ stars (Figs.~\ref{figure:abSurf2.0} and \ref{figure:abSurf2.5}), the abundances obtained with the various $g_\mathrm{rad}$ calculation methods agree within the observational uncertainty. The one exception is Ca for the $2.5~{M_\odot}$ wind model just before its evolution stopped: too high an abundance of Fe in the Z-bump caused the test atom approximation we used to be questionable. For these two masses, the overall agreement between OP and SVP2020 is slightly better than with SVP2004 for the RMT models, whereas SVP2004 is closer to the OP results for the wind models.
The spread in abundance is much smaller than for the $1.5~{M_\odot}$ models, especially for the models with RMT mixing. This is related to the smaller $g_\mathrm{rad}$ differences but also the more similar shapes of $g_\mathrm{rad}$ versus depth where chemical separation occurs. 

\begin{figure*}
   \centering
    \includegraphics[width=\textwidth]{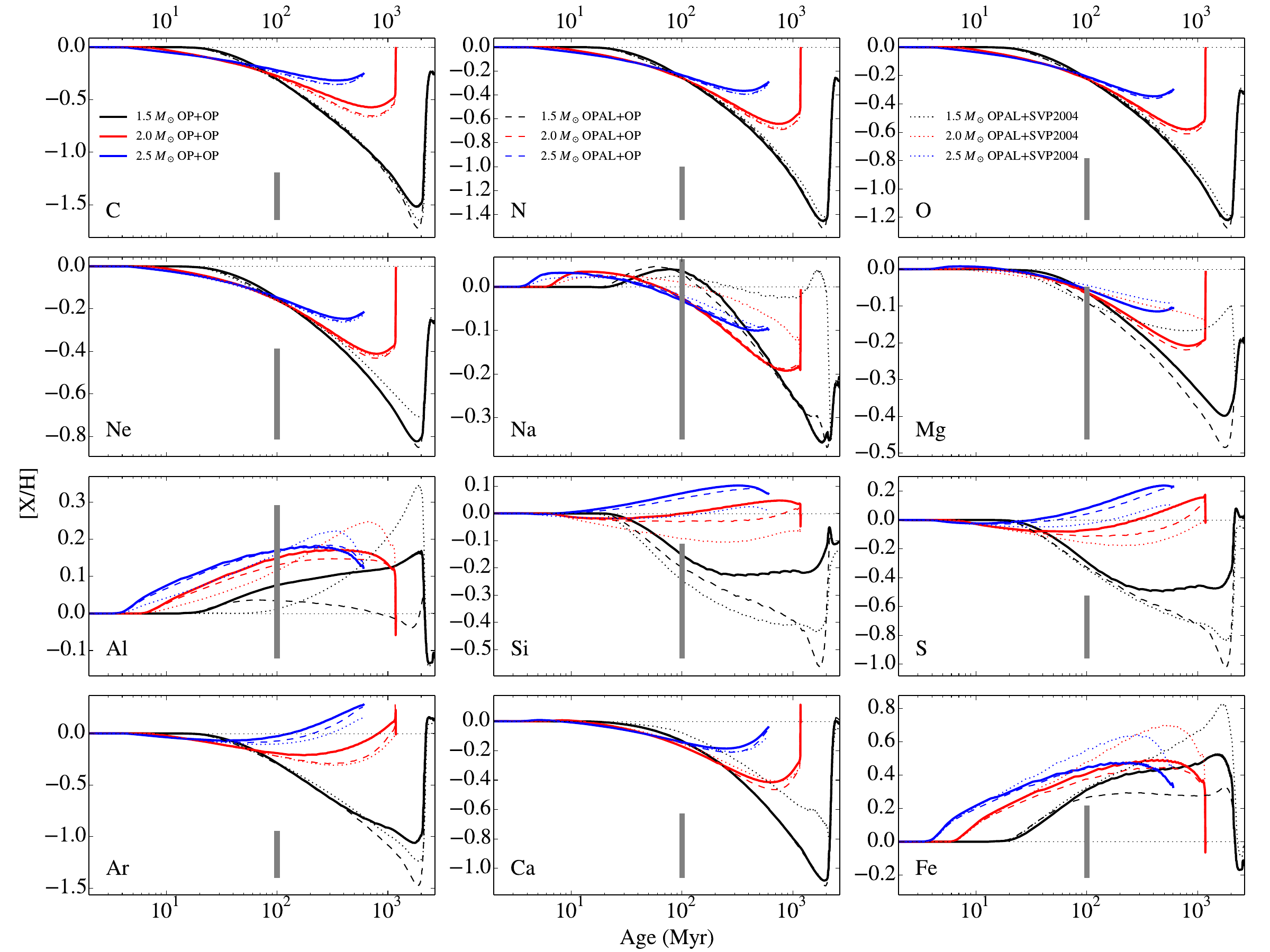}
      \caption{Abundance evolutions obtained with OP and OPAL opacity calculations for the three masses studied (1.5, 2, and 2.5~$M_\odot$, represented with black, red, and blue lines, respectively). The reference model used OP opacities along with OP $g_\mathrm{rad}$ (thick solid lines, `OP+OP'). The runs using OPAL opacities and OP or SVP2004  $g_\mathrm{rad}$ are shown with dashed (`OPAL+OP') or dotted (`OPAL+SVP2004') lines, respectively. The thin dotted line and the grey bar have the same meaning as in Figs.~\ref{figure:abSurf1.5} to \ref{figure:abSurf2.5}.
              }
         \label{figure:compareOPAL_OP}
\end{figure*}


\section{Discussion and conclusions}\label{discussion}

For the Am star models investigated, the three methods we used yield surface abundances that agree with each other, except for some elements at 1.5~$M_\odot$ when mass loss is considered. Since the SVP method has been developed to save computing time, we compared the durations of our different runs, summarised in Table~\ref{table:tempsCalcul}. The OP models were run with 16 cores (`OP' in Table~\ref{table:tempsCalcul}). The computing time uncertainty has been evaluated to be 10\%. Tests were done with $g_\mathrm{rad}$ computations in a sequential way for the RMT mixing models (`OP seq.' in Table~\ref{table:tempsCalcul}), showing that the parallelisation reduces the computing time by a factor of 3 to 4. These values are small compared to the number of cores used because a small fraction of the global computing time is used to evaluate $g_\mathrm{rad}$. We did not perform this test for the wind models owing to the duration of their computations. Little time (a few percent) is saved with the SVP runs compared to the OP parallel calculations, whereas we obviously gain the factor due to the change from parallel to sequential $g_\mathrm{rad}$ computations when compared to sequential OP runs. The difference between the time used with SVP2004 and SVP2020 is related to a slightly different number of diffusion steps. For the wind models, using the SVP method was valuable for the 1.5~$M_\odot$ model, with runs almost ten times shorter than the OP parallel run. This is due to the numerous diffusion steps -- and therefore $g_\mathrm{rad}$ calculations -- between two structural convergences for this mass. For higher masses, the evolution time steps are small and the convergence of the stellar structure dominates the global computing time, explaining the small gain in computing time with SVP.

\begin{table*}
\begin{center}
\tiny{
\begin{tabular}{ccccccccc}
\hline
Type&Model ($M_\odot$)&Comp. time (OP)&Comp. time (OP seq.)&OPseq./OP&SVP2020/OP&SVP2004/OP&SVP2020/OP seq.&SVP2004/OP seq.\\
\hline
{\multirow{3}{*}{RMT$\left.\vphantom{\begin{tabular}{l}l\\l\end{tabular}}\right\{$}}
&1.5&1h 18mn 31s&5h 31mn 54s&4.23&0.89&0.79&0.21&0.19\\
&2.0&1h 11mn 4s&3h 36mn 42s&3.0&0.92&0.87&0.30&0.28\\
&2.5&0h 44mn 23s&2h 52mn 52s&3.89&-&0.76&-&0.19\\
\hline
{\multirow{3}{*}{Wind$\left.\vphantom{\begin{tabular}{l}l\\l\end{tabular}}\right\{$}}
&1.5&14h 1mn 41s&-&-&0.11&0.11&-&-\\
&2.0&4d 4h 10mn 29s&-&-&0.70&0.92&-&-\\
&2.5&8d 1h 45mn 10s&-&-&-&1.03&-&-\\
\hline
\hline
\end{tabular}
}
\end{center}
\caption{Computing time of the OP $g_\mathrm{rad}$ parallel and sequential runs (`OP' and `OP seq.', respectively), and of the sequential OP, SVP2020, and SVP2004 runs scaled to the duration of the parallel OP run (`OP/OP seq.', `SVP20xx/OP'; xx=20 or 04 for SVP2020 and SVP2004, respectively) for each kind of model. Likewise, the last two columns show the ratios of the SVP runs to the sequential OP calculations. `Type' refers to the kind of model (`RMT' and `Wind' for RMT mixing and wind models, respectively). As wind model evolutions are very long, no sequential OP run was attempted. The SVP2020 wind model could not be evolved to the end of the MS and has been ignored.}\label{table:tempsCalcul}
\end{table*}

The computing time for models with radiative diffusion is mostly taken up by the $g_\mathrm{rad}$ calculations and those of the Rosseland means when calculated from monochromatic data. The opacity computations required to build the stellar structure dominate the time budget unless there are numerous diffusion steps between two evolution steps. In these cases, SVP approximations can reduce the duration of a run somewhat, especially when the OP $g_\mathrm{rad}$ are computed sequentially. In our tests, only the OP $g_\mathrm{rad}$ evaluations were performed sequentially, the Rosseland opacities still being computed in parallel. If Rosseland mean calculations are also performed in a sequential way, the gain obtained by using the SVP method would decrease because of the time used to evaluate the mean opacities, unless one uses tabulated mean opacities tables with chemical composition described as $X, Y,$ and $Z$, like the OPAL type 1 data; the drawback would be a loss of consistency between the local chemical composition and the mean opacity values. However, a test with the RMT mixing models shows that the differences between the abundances obtained with the OPAL opacities and those computed with the OP data remain within the observational error, except for S and Ar with OP, and Ca with SVP2004 in the $1.5~M_\odot$ model. Figure~\ref{figure:compareOPAL_OP} presents these results for models  with OPAL opacities and $g_\mathrm{rad}$ computed with either OP or SVP2004, in comparison with the reference runs with OP $g_\mathrm{rad}$ and opacities. We chose to only use the SVP2004 version due to its better agreement with the OP calculations for the $1.5~M_\odot$ stars. Concerning the MS evolution, the effective temperature ($T_\mathrm{eff}$) differences at a given luminosity are at most 30~K between the runs with OP mean opacities and $g_\mathrm{rad}$ and the OPAL calculations, either with OP or SVP2004 $g_\mathrm{rad}$, whatever the stellar mass. The $T_\mathrm{eff}$ determination uncertainty being $\pm200$~K for this temperature range \citep{Napiwotzki_etal1993}, the MS evolutions from the OPAL and OP opacity runs were considered to be the same. Table~\ref{table:tempsCalculOPAL} summarises the computing time using OPAL opacities and sequential OP or SVP2004 $g_\mathrm{rad}$, with the OP opacity and sequential OP $g_\mathrm{rad}$ run in comparison, which is considered the reference run. When using SVP2004, the runs are 10 to more than 20 times shorter. The SVP approximation with OPAL opacities can now be used to compute large grids of models that could be used to bracket the stellar fundamental parameters. A finer grid could then be calculated using OP for $g_\mathrm{rad}$ as well as for the Rosseland means to refine their determination.

\begin{table*}
\begin{center}
\tiny{
\begin{tabular}{ccccc}
\hline
Model ($M_\odot$)&Comp. time (OPAL+OP)&Comp. time (OPAL+SVP2004)&OPAL+OP/Ref.&OPAL+SVP2004/Ref.\\
\hline
1.5&4h 46mn 24s&15mn 47s&0.86&0.05\\
2.0&2h 19mn 31s&22mn 45s&0.64&0.10\\
2.5&1h 12mn 38s&7mn 36s  &0.42&0.04\\
\hline
\hline
\end{tabular}
}
\end{center}
\caption{Computing time of the OPAL opacity RMT mixing runs with $g_\mathrm{rad}$ computed with OP and SVP2004 (`OPAL+OP' and `OPAL+SVP2004', respectively), and corresponding ratios vs the reference run (`Ref.') with $g_\mathrm{rad}$ computed with OP sequentially and OP opacities.}
\label{table:tempsCalculOPAL}
\end{table*}

As \cite{Deal_etal2018} show that radiative diffusion has detectable effects on the evolution and oscillation properties of solar-type pulsators, further studies could address this kind of star and check if the two-step stellar grid building can be used considering the asteroseismic constraints. If so, this could be useful owing to the measurement precision reached and the number of targets observed with present (e.g. {\it Kepler}, \citealt[][]{Gilliland_etal2010}, or TESS, \citealt{Ricker_etal2015}) and future instruments (PLATO; \citealt{Rauer_etal2014}). Constraints on the mixing processes of F stars could also be revisited \citep{Deal_etal2020,Deal_etal2023}. An investigation of higher masses would also be interesting since radiative diffusion could have an impact on their oscillations, especially for hybrid pulsators \citep{Hui-Bon-Hoa_Vauclair2018b,Hui-Bon-Hoa_Vauclair2018a} and slowly pulsating B stars \citep{Rehm_etal2024}.

\begin{acknowledgements}
I thank the anonymous referee for questions and remarks that helped improve the manuscript. I am also grateful to G. Alecian for the discussion we had about the SVP method and his careful reading of the manuscript. My thanks also go to F. LeBlanc for having taken time to suggest corrections to the first draft.
\end{acknowledgements}

%
%

   \bibliographystyle{aa} 
   \bibliography{G_radAmFm}

\end{document}